\documentclass{jfm}

\usepackage{graphicx}
\usepackage{natbib}
\usepackage{amsmath}
\usepackage{psfrag}
\usepackage{psfragx}
\usepackage{epsfig}
\usepackage{xcolor,import}
\usepackage{pstool}
\usepackage{placeins}
\usepackage{amssymb}
\usepackage{mathabx}
\usepackage{graphicx,xcolor}        
\usepackage{subfig}                             
\usepackage{wasysym}        
\usepackage{floatrow}
\usepackage{dashrule}
\usepackage{tikz}
\usetikzlibrary{arrows,decorations.markings}
\usetikzlibrary{calc}
\ifCUPmtlplainloaded \else
  \checkfont{eurm10}
  \iffontfound
    \IfFileExists{upmath.sty}
      {\typeout{^^JFound AMS Euler Roman fonts on the system,
                   using the 'upmath' package.^^J}%
       \usepackage{upmath}}
      {\typeout{^^JFound AMS Euler Roman fonts on the system, but you
                   dont seem to have the}%
       \typeout{'upmath' package installed. JFM.cls can take advantage
                 of these fonts,^^Jif you use 'upmath' package.^^J}%
      }
  \else
  \fi
\fi

\ifCUPmtlplainloaded \else
  \checkfont{msam10}
  \iffontfound
    \IfFileExists{amssymb.sty}
      {\typeout{^^JFound AMS Symbol fonts on the system, using the
                'amssymb' package.^^J}%
       \usepackage{amssymb}%

      }{}
  \fi
\fi

\ifCUPmtlplainloaded \else
  \IfFileExists{amsbsy.sty}
    {\typeout{^^JFound the 'amsbsy' package on the system, using it.^^J}%
     \usepackage{amsbsy}}
    {\providecommand\boldsymbol[1]{\mbox{\boldmath $##1$}}}
\fi

\newcommand\Rey{\mbox{\textit{Re}}}  

\newsavebox{\astrutbox}
\sbox{\astrutbox}{\rule[-5pt]{0pt}{20pt}}

\title[]{Two-point similarity in\\
  the round jet revisited}

\author[Azur Hod\v zi\' c and Clara M. Velte]%
{Azur Hod\v zi\' c$^1$%
  \thanks{Email address for correspondence: azuhod@mek.dtu.dk}\ns
and Clara M. Velte$^1$}

\affiliation{$^1$Department of Mechanical Engineering, Technical University of Denmark,
2800, Kgs. Lyngby, Denmark\\[\affilskip]}

\pubyear{2010}
\volume{650}
\pagerange{119--126}
\date{?; revised ?; accepted ?. - To be entered by editorial office}
\begin{document}

\maketitle

\begin{abstract}
The similarity of the two-point correlation tensor along the streamwise direction in the axi-symmetric jet far-field is analyzed, herein its utility in spectral theory. A separable two-point correlation coefficient has been the basis for the argument that the energy-optimized basis functions along the streamwise direction are Fourier modes (from the approach of equilibrium similarity theory). This would naturally be highly desirable both from a computational and an analytical perspective. The present work, however, shows that the two-point correlation tensor multiplied by the Jacobian is not displacement invariant even in logarithmically stretched coordinates. This result directly impacts the motivation for a Fourier-based representation of the correlation function in spectral space in relation to the Proper Orthogonal Decomposition (POD) of the field. It is demonstrated that a displacement invariant form of the kernel is impossible to achieve using the suggested coordinate transformations from earlier works. This inability is shown to be related to the fundamental differences between the turbulent flow at hand and the ideal case of homogeneous turbulence.
\end{abstract}
\section{Introduction}
The search for a similarity solution is essentially a search for an optimal coordinate system. In combination with corresponding scaling laws a similarity solution manifests itself in terms of a collapse of the statistical variants (mean velocity components, Re-stresses etc.) in the respective coordinate system.

Historically, self-similar turbulent flows (e.g. jets and wakes) have been analyzed by introducing the similarity coordinate $\eta=r/\delta\left(x\right)$, \cite{Johansson2003}, \cite{Johansson2006} along which the scaled statistical variants collapse. The exact definition of the coordinate system including the determination of $\delta(x)$ is deduced directly from the two-point correlation equations by hypothesizing that the flow has reached a state of self-similarity. This implies that the inherent kinematic processes of the flow develop in a self-similar way with the downstream coordinate \cite{Ewing1995}, \cite{Ewing2007}, \cite{Nedic2013a}. Self-similar features in jets have been recognized in earlier works by demonstrating the collapse of velocity profiles at various streamwise coordinates - and even more ambitiously in the collapse of two-point correlation coefficients \cite{Ewing2007}, \cite{Wanstrom2009} of the scaled velocity field. 

The observed collapse of the two-point correlation coefficients in \cite{Ewing2007} was used as a quantification of the self-similarity of scales at different absolute distances downstream from the nozzle by utilizing a logarithmic stretching of the grid. The collapse led to the conclusion that the modes could be optimally represented in terms of energy by a Fourier basis. As a consequence, it was concluded that the jet could be homogenized in the streamwise direction by the proposed coordinate transformation. If true this would lead to significant simplifications both analytically and computationally for the decomposition of the far-field and would essentially allow a space-time decomposition of the far-field region of the jet, where an analytical Fourier basis could be used in three out of four (space and time) coordinate directions. 

The current work revises the implications of the collapse of the two-point correlation coefficient in lieu of its role in spectral theory - herein the Proper Orthogonal Decomposition (POD). Particularly the conclusions drawn in \cite{Ewing2007} related to this collapse are revised from the perspective of the POD integral. This revision is imperative for the representation of the round jet far-field in spectral space, its decomposition using analytical basis functions, and not least for the physical understanding of the far-field turbulence of the round jet.
%
%
%
\section{Experimental procedure\label{sec:experimental_procedure}}
Two-component Particle Image Velocimetry (PIV) measurements were performed in the far-field region of a turbulent axi-symmetric jet in a tent of dimensions ${2.5\times3.0\times10.0\,\mathrm{m^3}}$. The tent was seeded with glycerin droplets of approximately ${2-3\,\mathrm{\mu m}}$ using an in-house seeding generator equipped with an atomizing Laskin nozzle. 
\begin{figure}[h]%
\centering  
\def\svgwidth{0.6\textwidth} 
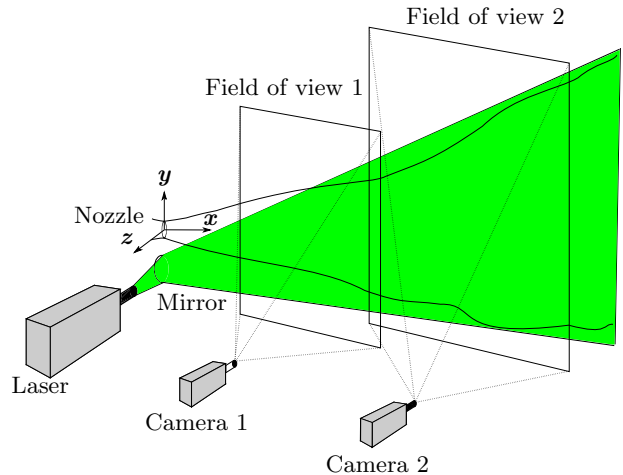  
\caption{The experimental setup consisting of a laser reflecting of a mirror and two PIV cameras used to extend the total field of view.\label{fig:experimental_setup}}
\end{figure}
\FloatBarrier
\noindent
The jet was driven by a fan located inside the tent ensuring a homogeneous seeding distribution. The axi-symmetrical nozzle design was based on fifth-order polynomials in order to create a smooth contraction ratio from, $60\,\mathrm{mm}$ to~${D=10\,\mathrm{mm}}$. The nozzle was mounted on the same jet box used in~\cite{Gamard2004}, \cite{Ewing2007} and in \cite{Wanstrom2009}.

The experimental setup consisted of two $16\,\mathrm{MPix}$ ($4872\times3248\,\mathrm{pix}$) Dantec FlowSenseEO cameras, with a pixel pitch of $7.4\,\mathrm{\mu m}$ using $60\, \mathrm{mm}$ Nikon lenses with an aperture of $f^{\#}\,2.8$. A dual cavity $200\,\mathrm{mJ}$ ND:YAG $532\,\mathrm{nm}$ laser was used to illuminate the particles. A sketch of the experimental configuration is shown in figure \ref{fig:experimental_setup}. 

The combined field-of-view (FOV) spanned from $33.1D$ to $108.1D$ from the nozzle. The time between pulses was carefully optimized in order to achieve a suitable balance between dynamic range and pixel locking, where the aim was to maximize the dynamic range, by reducing particle loss in the field-of-view (FOV) region of camera 1 closest to the nozzle and minimizing the peak-locking bias at the downstream region of the FOV of camera 2 (see figure \ref{fig:experimental_setup}). This resulted in a time between pulses of $150\,\mathrm{\mu s}$. 

A sampling rate of $1\,\mathrm{Hz}$ was used in the experiments in order to obtain $11\,000$ uncorrelated realizations, which were acquired in a single sitting. The experiment was performed at $\Rey=20\,000$ based on the nozzle diameter, $D=1\mathrm{cm}$, and nozzle exit velocity, $U_0=30\mathrm{m/s}$. 

\FloatBarrier
\section{Revision of two-point similarity}
\cite{Ewing2007} presented a two-point symmetry, based on the following definition of the two-point correlation coefficient along the streamwise direction in Cartesian coordinates, \cite{Hinze1975}
\begin{equation}
\rho_{uu}\left(x,x'\right) = \frac{\overline{u(x,t)u(x',t)}}{\sqrt{\overline{u(x)^2}\,\overline{u(x')^2}}},\label{eq:correlation_coefficients}
\end{equation}
where $u(x,t)$, and $u(x',t)$ are the fluctuating parts of the streamwise velocity components at two different points, $x$ and $x'$, along the streamwise direction of the jet far-field. The overline designates temporal averaging. The velocity field was sampled using an experimental setup consisting of a hot-wire probe and an LDA system in order to obtain a two-point temporally averaged correlation tensor along the centerline of the jet in the self-similar region (see \cite{Ewing2007} for the details of the experimental setup). 

\begin{center}
\begin{figure}[t]
\subfloat[]{\includegraphics[scale=0.45]{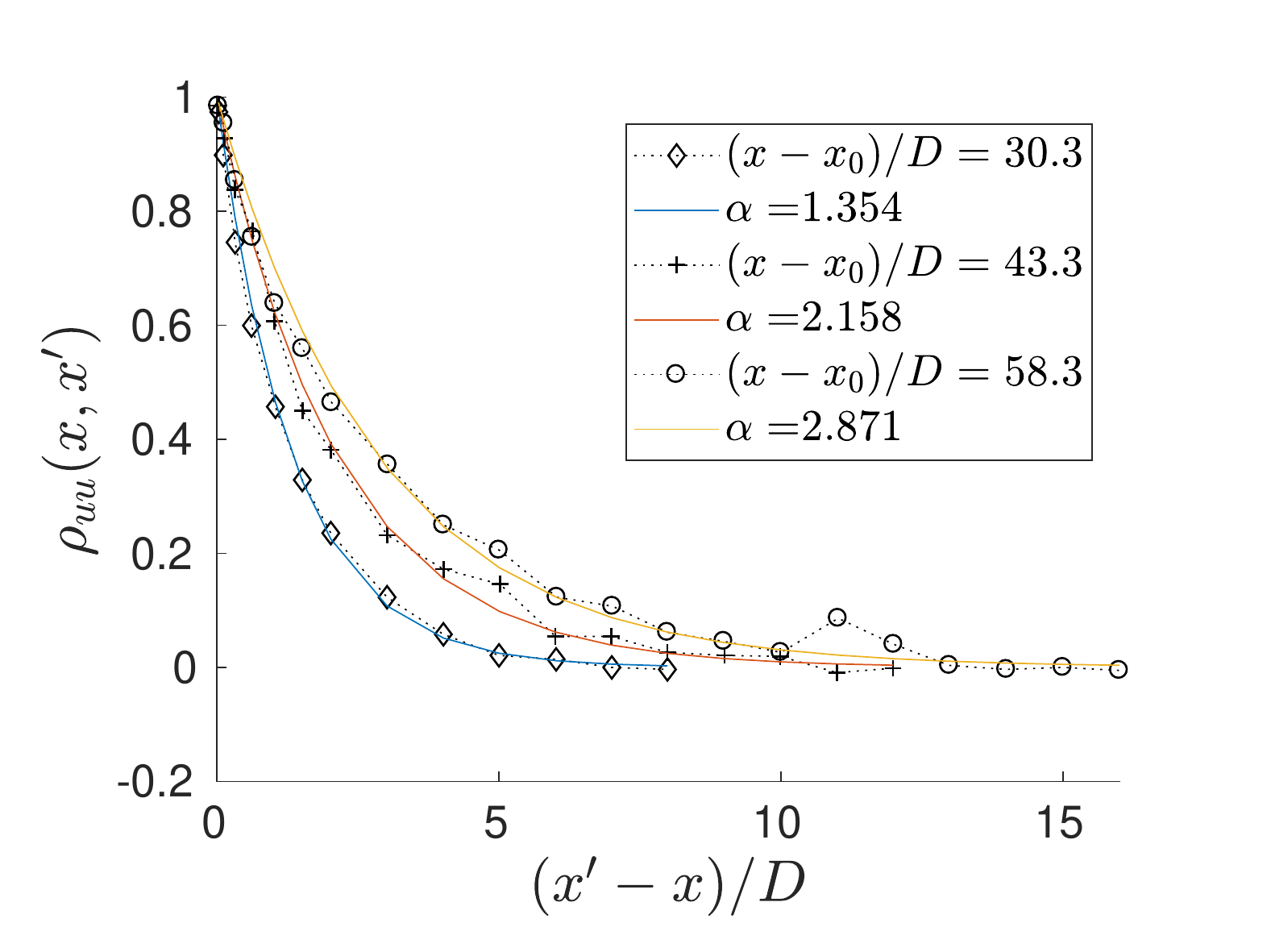}\label{fig:scaled_correlationtensor}}
\subfloat[]{\includegraphics[scale=0.45]{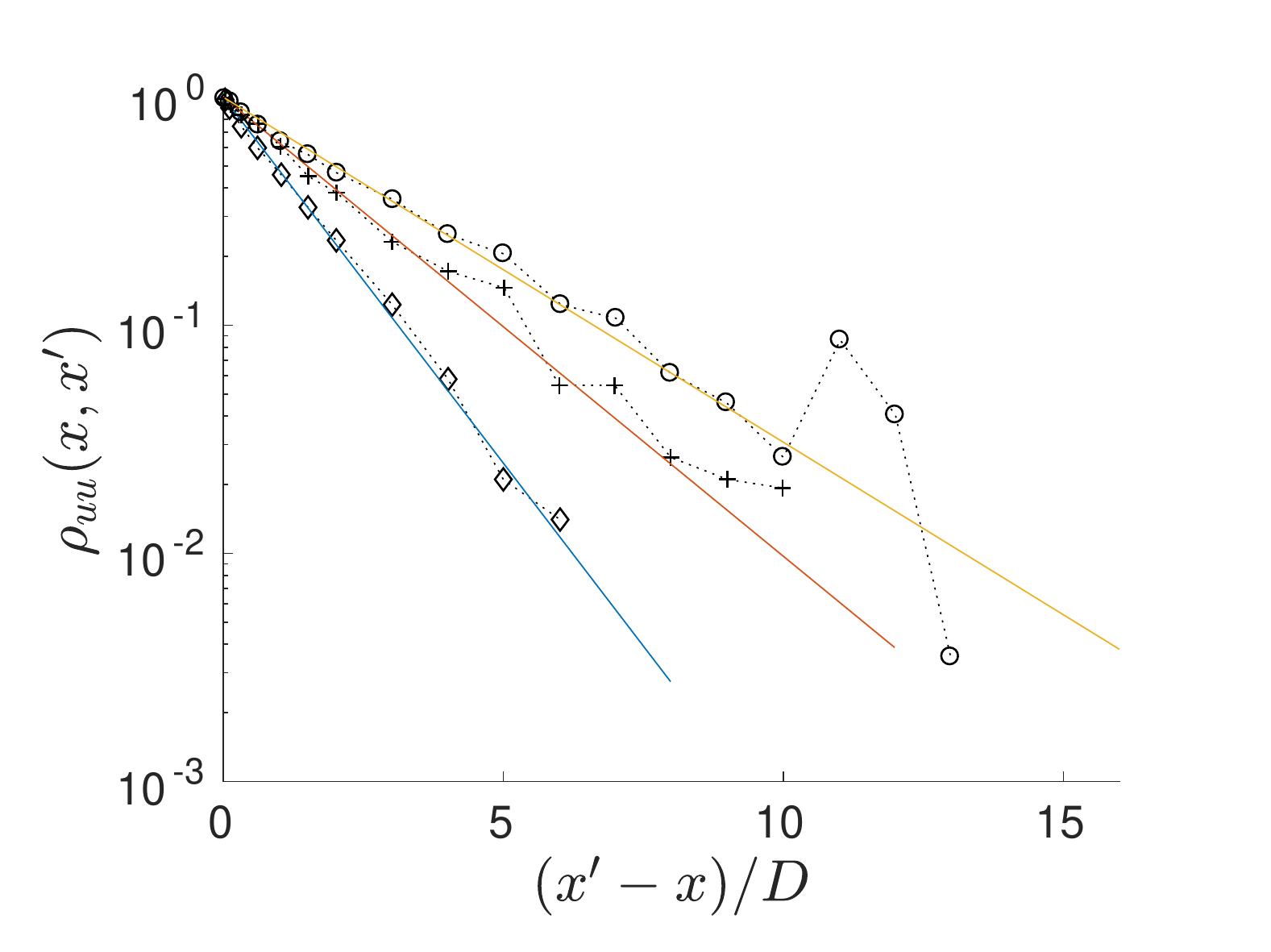}\label{fig:scaled_correlationtensor_log}}
\caption{Two-point correlation coefficient along the jet centerline from \cite{Ewing2007} where the full lines correspond to the analytical form \eqref{eq:analytical_form} for various $\alpha$-values. (a): linear representation, (b): semi-logarithmic representation illustrating the exponential decay.
\label{fig:scaled_correlationtensors}}
\end{figure}
\end{center}

\begin{center}
\begin{figure}[h]
\subfloat[]{\includegraphics[scale=0.45]{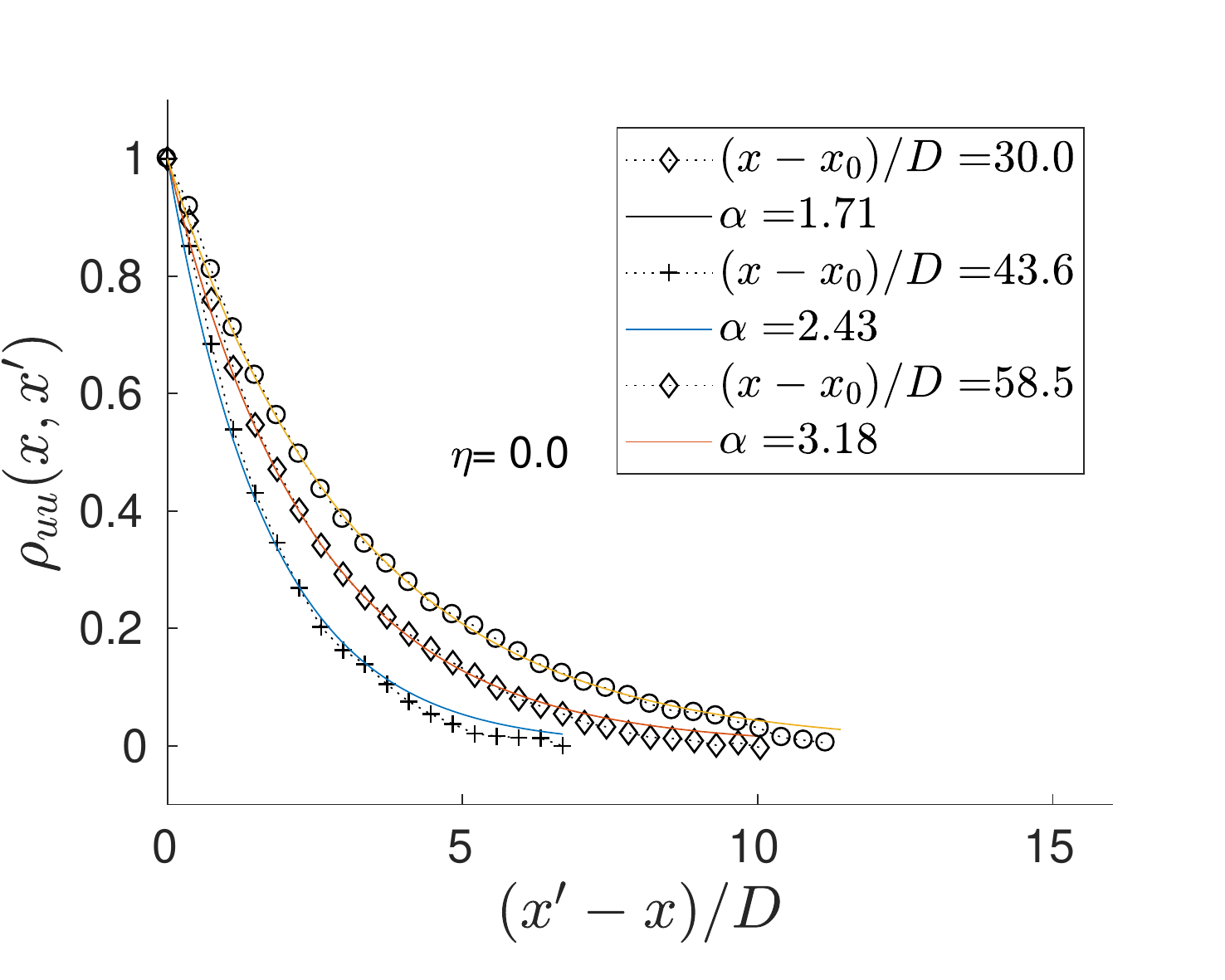}\label{fig:correlationtensor_new1}}
\subfloat[]{\includegraphics[scale=0.45]{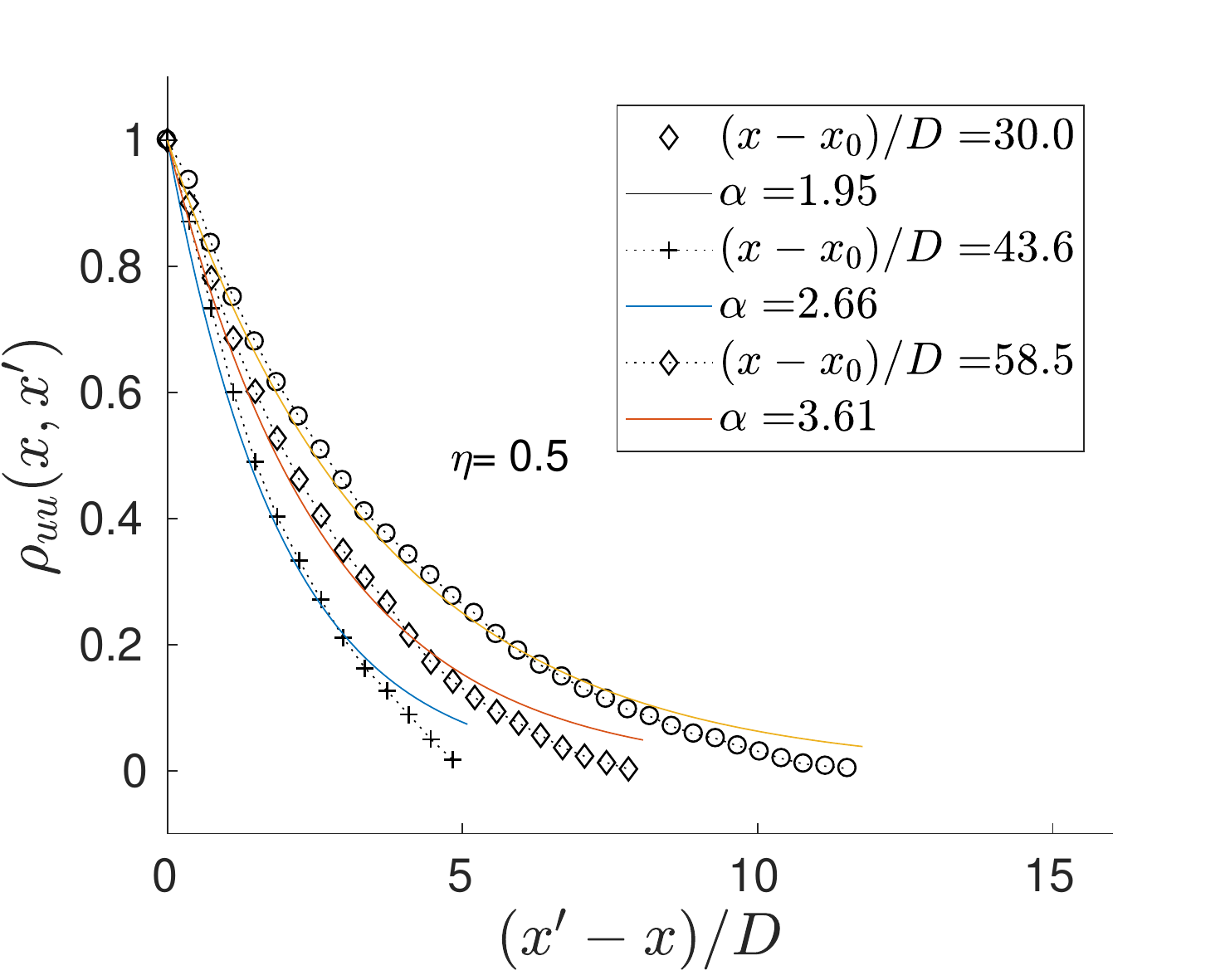}\label{fig:correlationtensor_new2}}\\
\subfloat[]{\includegraphics[scale=0.45]{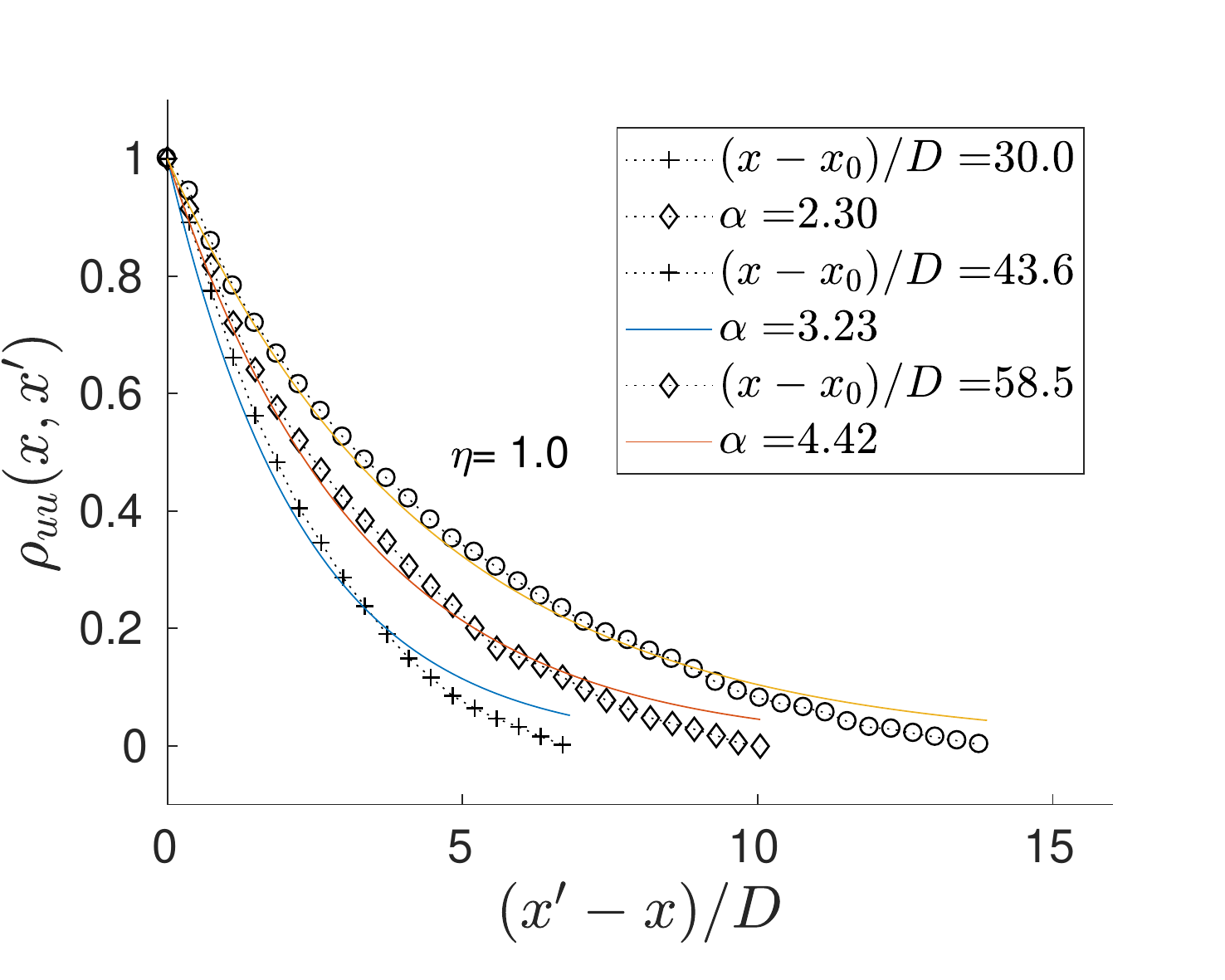}\label{fig:correlationtensor_new3}}
\subfloat[]{\includegraphics[scale=0.45]{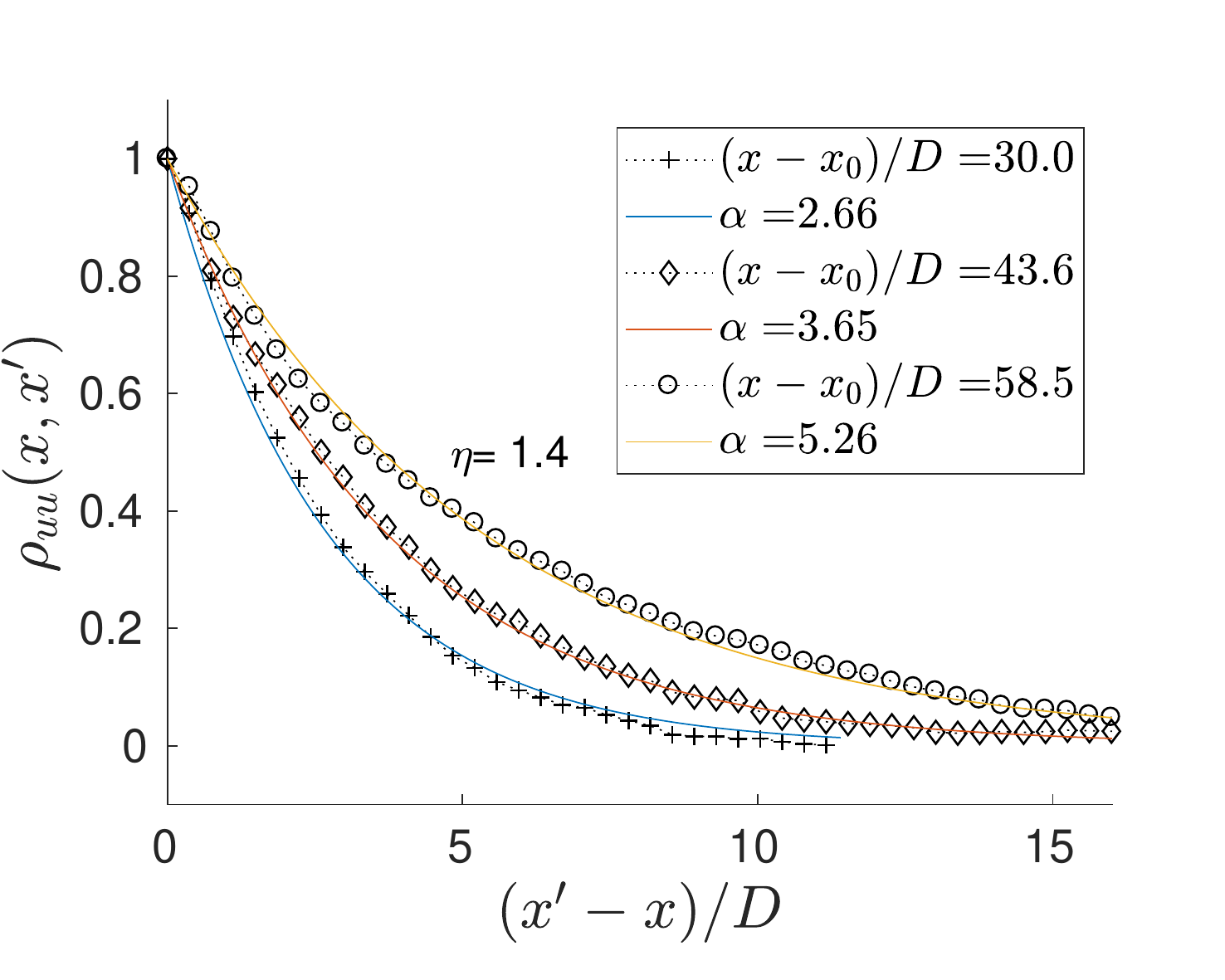}\label{fig:correlationtensor_new4}}\\
\subfloat[]{\includegraphics[scale=0.45]{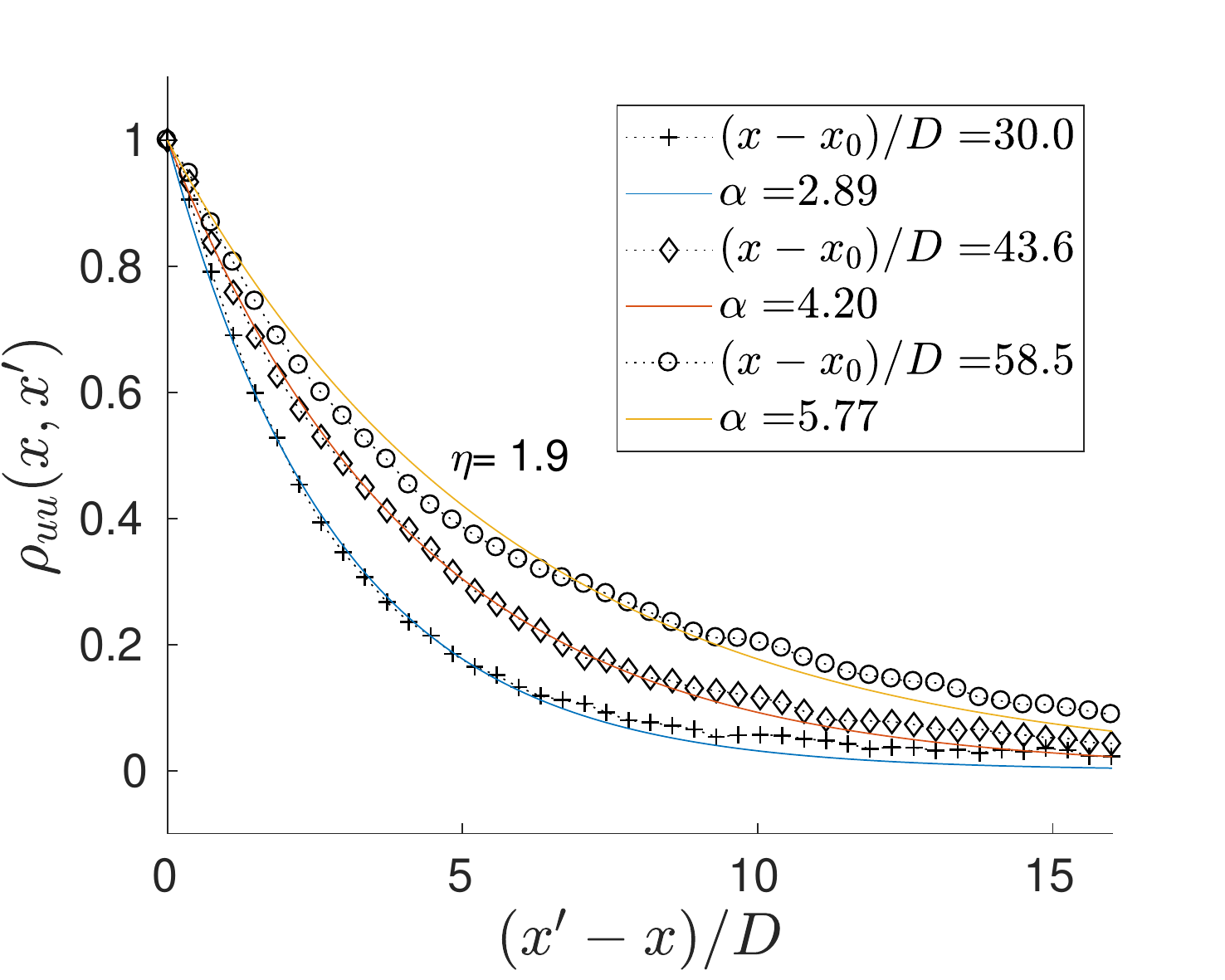}\label{fig:correlationtensor_new5}}
\subfloat[]{\includegraphics[scale=0.45]{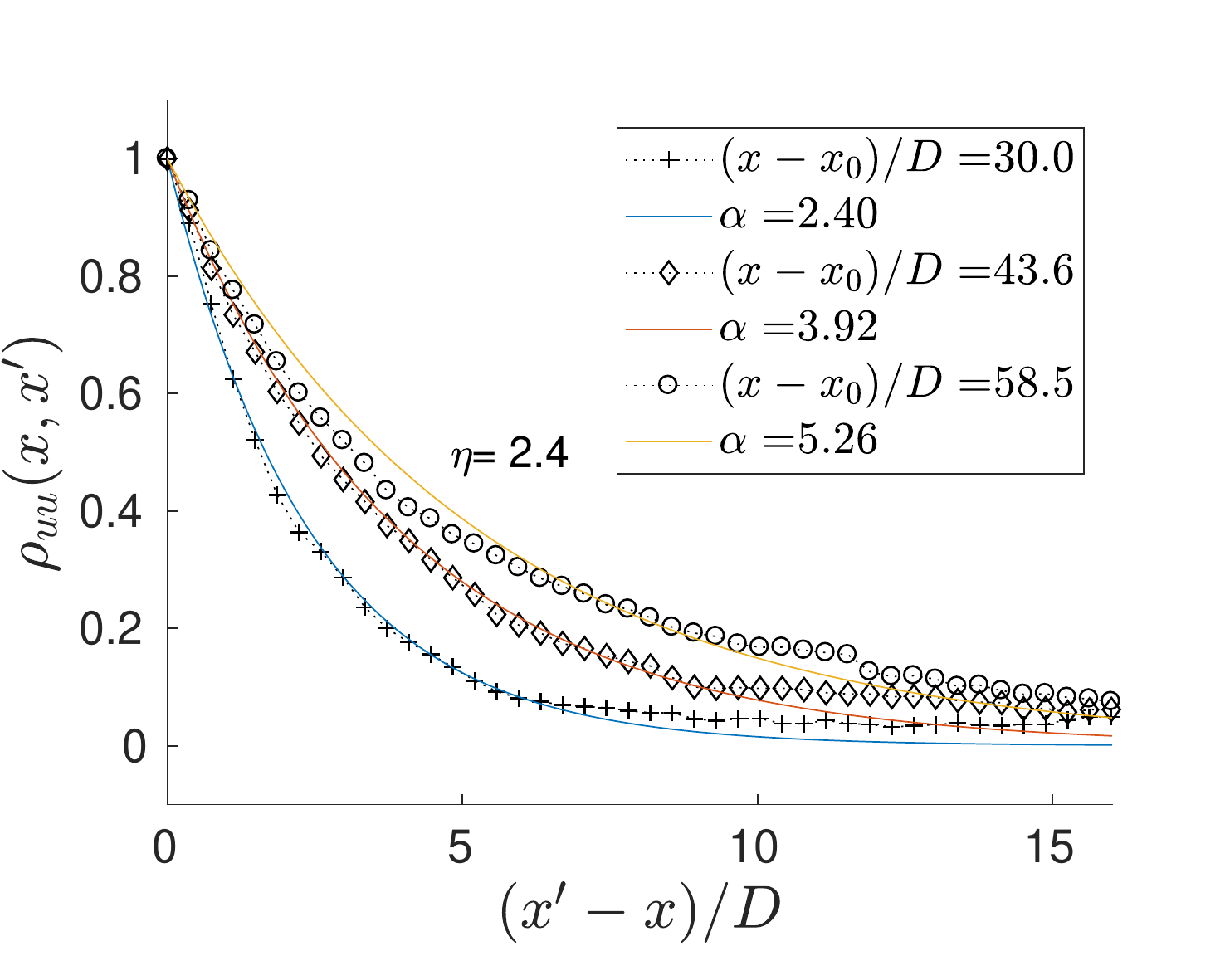}\label{fig:correlationtensor_new6}}
\caption{The development of $\rho_{uu}(x,x')$ with downstream position at transverse positions, ${\eta = [0,\,0.5,\,1.0,\,1.4,\,1.9,\,2.4]}$\label{fig:correlationtensors}.}
\end{figure}
\end{center}

The measured two-point correlation from \cite{Ewing2007} is seen in figure \ref{fig:scaled_correlationtensor} and was shown by \cite{Ewing2007} to be dependent on the offset position i.e. the correlation lengths (based on the correlation coefficient definition of \eqref{eq:correlation_coefficients}) are increasing downstream when sampled equidistantly. A two-point similarity hypothesis was then stated, which was hypothesized to admit to a separation of the two-point correlation function into a product of a function, $Q^{(i,j)}_s$, dependent of the absolute downstream position and a function, $q_{i,j}$, which is dependent on the streamwise logarithmic coordinate difference $\upsilon$, i.e.
\begin{equation}
\overline{u_i(x,y,z,t)u'_j(x',y',z',t)} = Q^{(i,j)}_s\left(x,x'\right)q_{i,j}\left(\upsilon,\eta,\eta',\theta,\theta',*\right),\label{eq:seperability_hypothesis}
\end{equation}
where
\begin{equation}
Q^{(i,j)}_s\left(x,x'\right) = a^{i,j}_c(*) U_s(x)U_s(x').
\end{equation}
The asterisk designates a dependency on initial conditions, \cite{George1989}, $a^{i,j}_c(*)$ are parameters that can depend on these conditions, \cite{Ewing2007}, and $U_s(x)$ is the local mean centerline velocity. The logarithmically stretched streamwise coordinate, $\xi$, is defined together with the transverse coordinate, $\eta$, and the azimuthal coordinate, $\theta$, in the following way
\FloatBarrier
\noindent
\begin{equation}
\xi = \ln\left(\frac{x-x_0}{D}\right)\hspace*{0.1cm},\hspace*{0.1cm} \eta = \frac{\sqrt{y^2+z^2}}{\delta(x)}\hspace*{0.1cm},\hspace*{0.1cm}\theta = \arctan\frac{z}{y},\label{eq:xi}
\end{equation}
such that the jet centerline is defined as $y=z=0$ in Cartesian coordinates (or as $\eta=0$ in the stretched coordinate). The jet half-width is defined as $\delta(x) = A(x-x_0)$, where $A=0.0926$ is the jet spreading rate and $x_0=3.1D$ is the virtual origin. The streamwise logarithmic separation coordinate is defined as
\begin{equation}
\upsilon = \xi'-\xi\label{eq:upsilon}.
\end{equation}
In order to characterize the two-point correlation coefficient and its downstream development, the results from \cite{Ewing2007} are shown in figure \ref{fig:scaled_correlationtensor} and \ref{fig:scaled_correlationtensor_log} together with the analytical form
\begin{equation}
\phi(x,x') = e^{-\frac{|x'-x|}{\alpha}},\label{eq:analytical_form}
\end{equation}
where the coefficients, $\alpha$, are assumed to be a function of $x$ in order to facilitate the modeling of the downstream development of the two-point correlation tensor, i.e. ${\alpha=\alpha(x)}$. The $\alpha$-values are summarized in figure \ref{fig:scaled_correlationtensor} for each downstream offset coordinate where the root-mean-square deviation between the data and \eqref{eq:analytical_form} relative to the $L^2$-norm of the data is $4.8\,\%$, $6.9\,\%$, and $8.8\,\%$ for the $x/D=30.3$, $43.3$, and $58.3$, respectively. Corresponding analytical forms have been displayed for the PIV data in figure \ref{fig:correlationtensors} where the fits are shown across the entire span of the jet, i.e. $\eta\in[0:2.4]$. The approximation $\phi(x,x')\approx\rho_{uu}(x,x')$ is useful as it allows us to evaluate, how the two-point coefficient develops downstream. This development is therefore modeled by the following
\begin{equation}
\frac{\overline{u(x,t)u(x',t)}}{\sqrt{\overline{u(x)^2}\,\overline{u(x')^2}}} \approx e^{-\frac{|x'-x|}{\alpha}} \label{eq:approx},
\end{equation}
where the two-point correlation of streamwise Reynolds stresses can be replaced by the scaling of \cite{Ewing2007}
\FloatBarrier
\noindent
\begin{center}
\begin{figure}[h]
\subfloat[]{\includegraphics[scale=0.45]{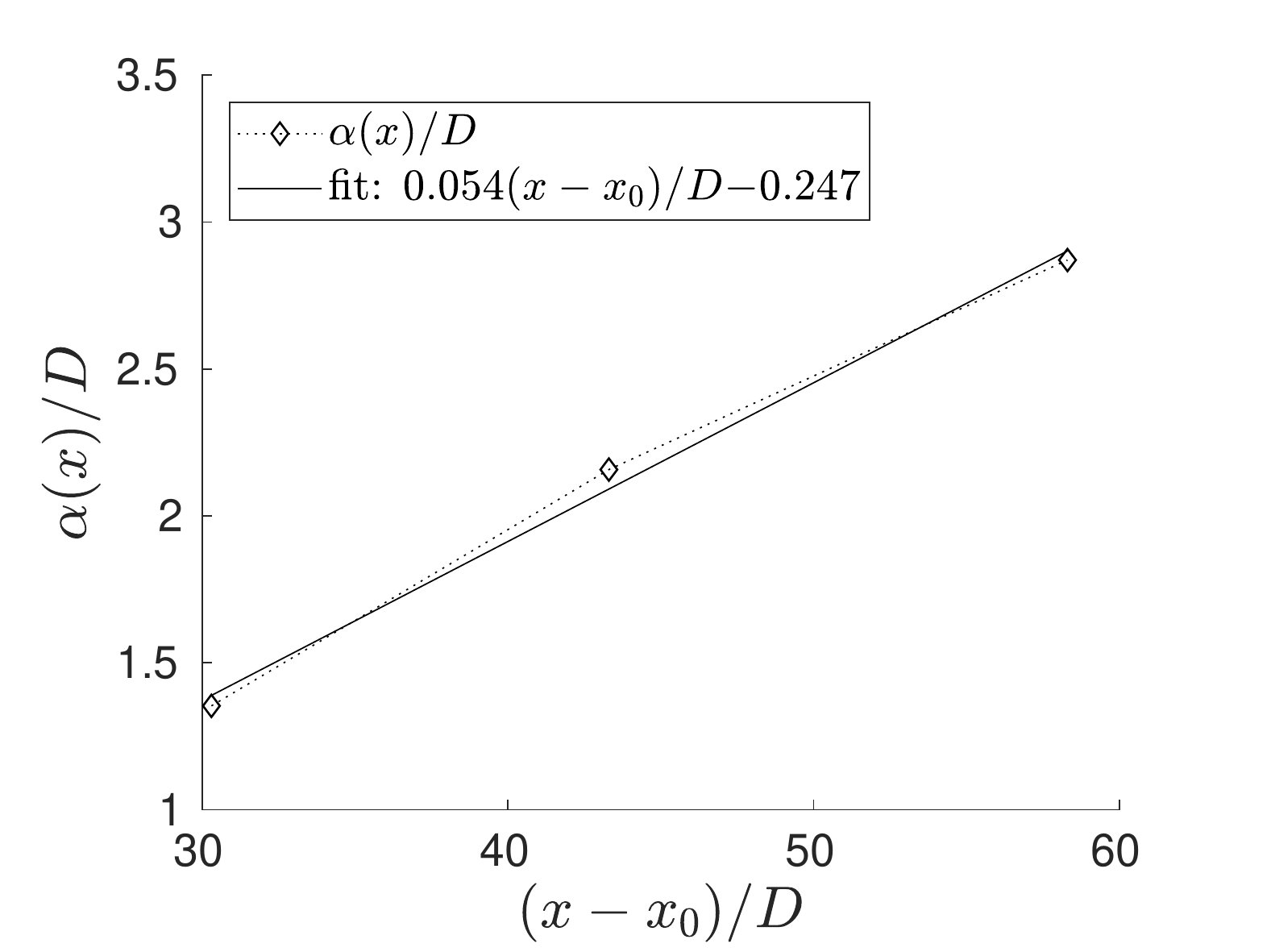}}
\subfloat[]{\includegraphics[scale=0.45]{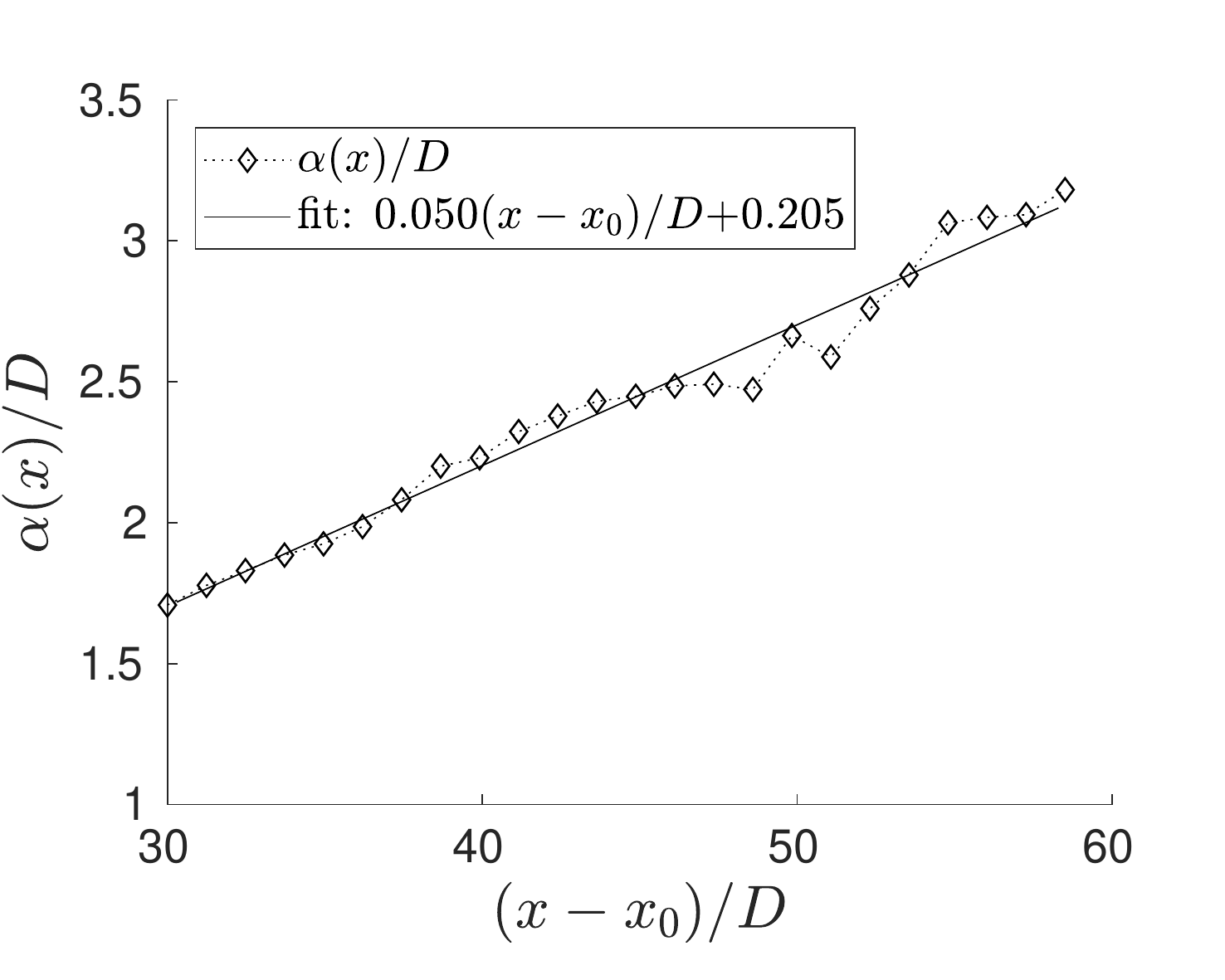}}
\caption{(a): Exponential coefficients, $\alpha(x)$, on the centerline obtained from a least squares fit to $\rho_{uu}(x,x')$ using data from \cite{Ewing2007} together with a linear fit, (b): corresponding results from PIV data.\label{fig:alpha}}
\end{figure}
\end{center}
\begin{equation}
\sqrt{\overline{u(x)^2}\,\overline{u(x')^2}}=a^{1,1}_c(*) U_s(x)U_s(x').
\end{equation}
This leads to the following form describing the development of the two-point correlation function scaled by the mean centerline velocities
\begin{equation}
\frac{\overline{u(x,t)u(x',t)}}{a^{1,1}_c(*) U_s(x)U_s(x')} \approx e^{-\frac{|x'-x|}{\alpha}}.\label{eq:exp_alpha}
\end{equation}
Effectively, the results in \cite{Ewing2007} mean that the left-hand-side should be independent of absolute coordinates as it should only depend on coordinate differences in the streamwise $x$-direction i.e.
\begin{equation}
\frac{\overline{u(x,t)u(x',t)}}{a^{1,1}_c(*) U_s(x)U_s(x')} = q_{1,1}(\upsilon)\approx e^{-\frac{|x'-x|}{\alpha}}.
\end{equation}
From figure \ref{fig:scaled_correlationtensor_log} it is seen that $\alpha$ can be estimated by a linear form
\begin{equation}
\alpha(x) = ax+b,\label{eq:linear_fit}
\end{equation}
where the $x$-coordinate represents the offset of the correlation function. Figure \ref{fig:alpha} shows the $\alpha$-values together with \eqref{eq:linear_fit} for both the data of \cite{Ewing2007} and the current PIV data.
\FloatBarrier
\subsection{The relation between $\alpha(x)$ and the integral length scale}
\begin{center}
\begin{figure}[h]
\subfloat[]{\includegraphics[scale=0.45]{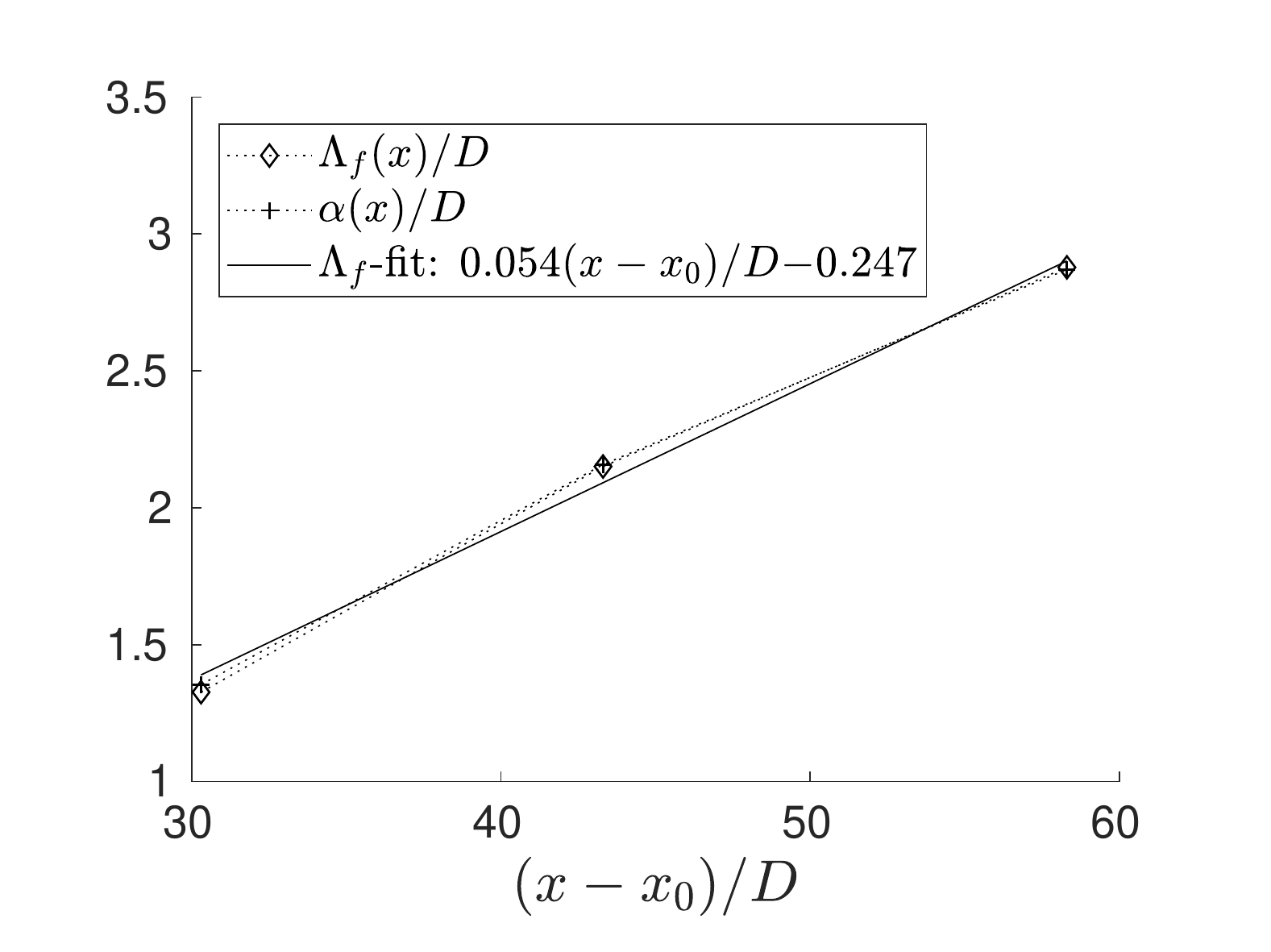}\label{fig:Lambda_alpha}}
\subfloat[]{\includegraphics[scale=0.45]{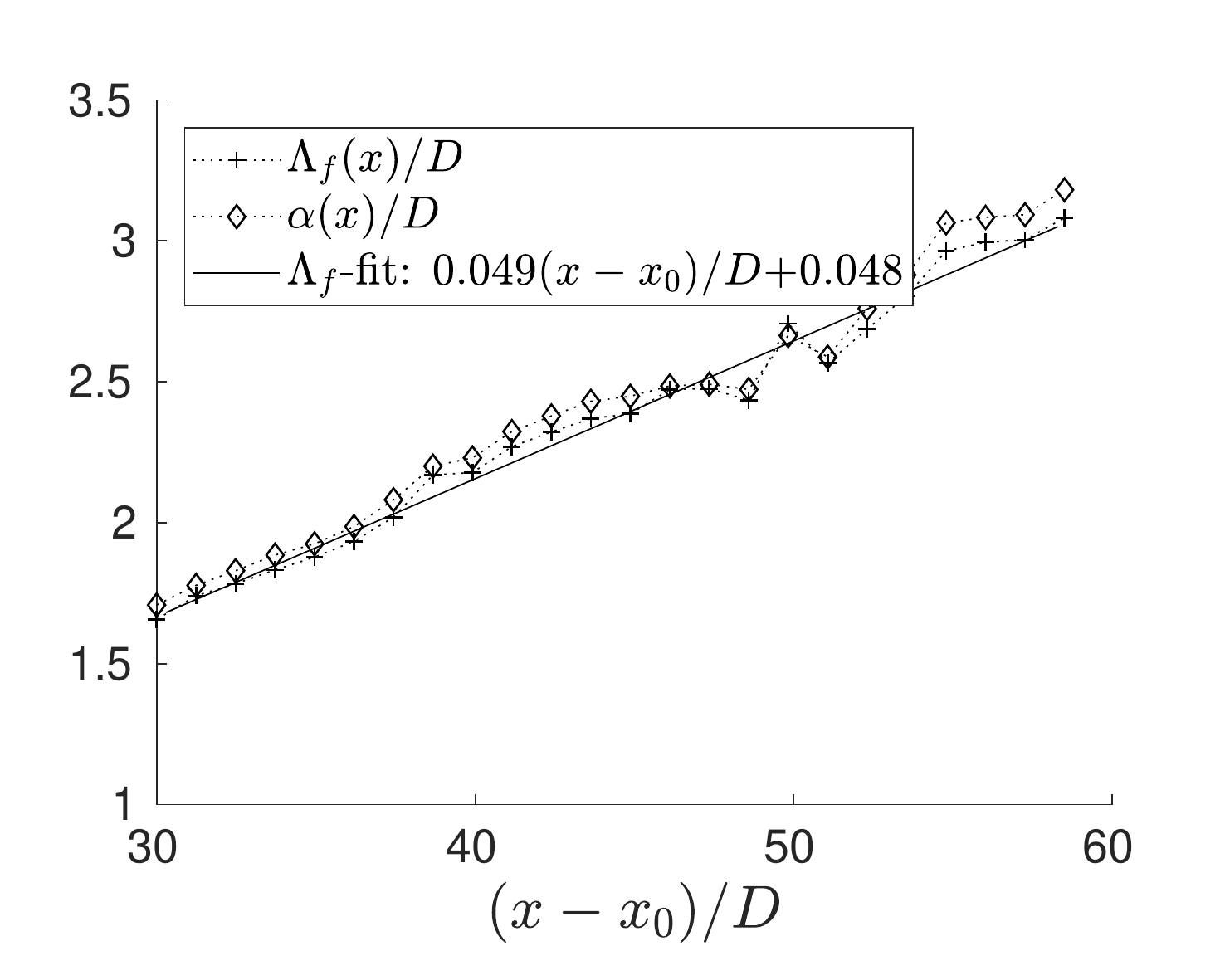}\label{fig:Lambda_alpha_new}}\\
\subfloat[]{\includegraphics[scale=0.45]{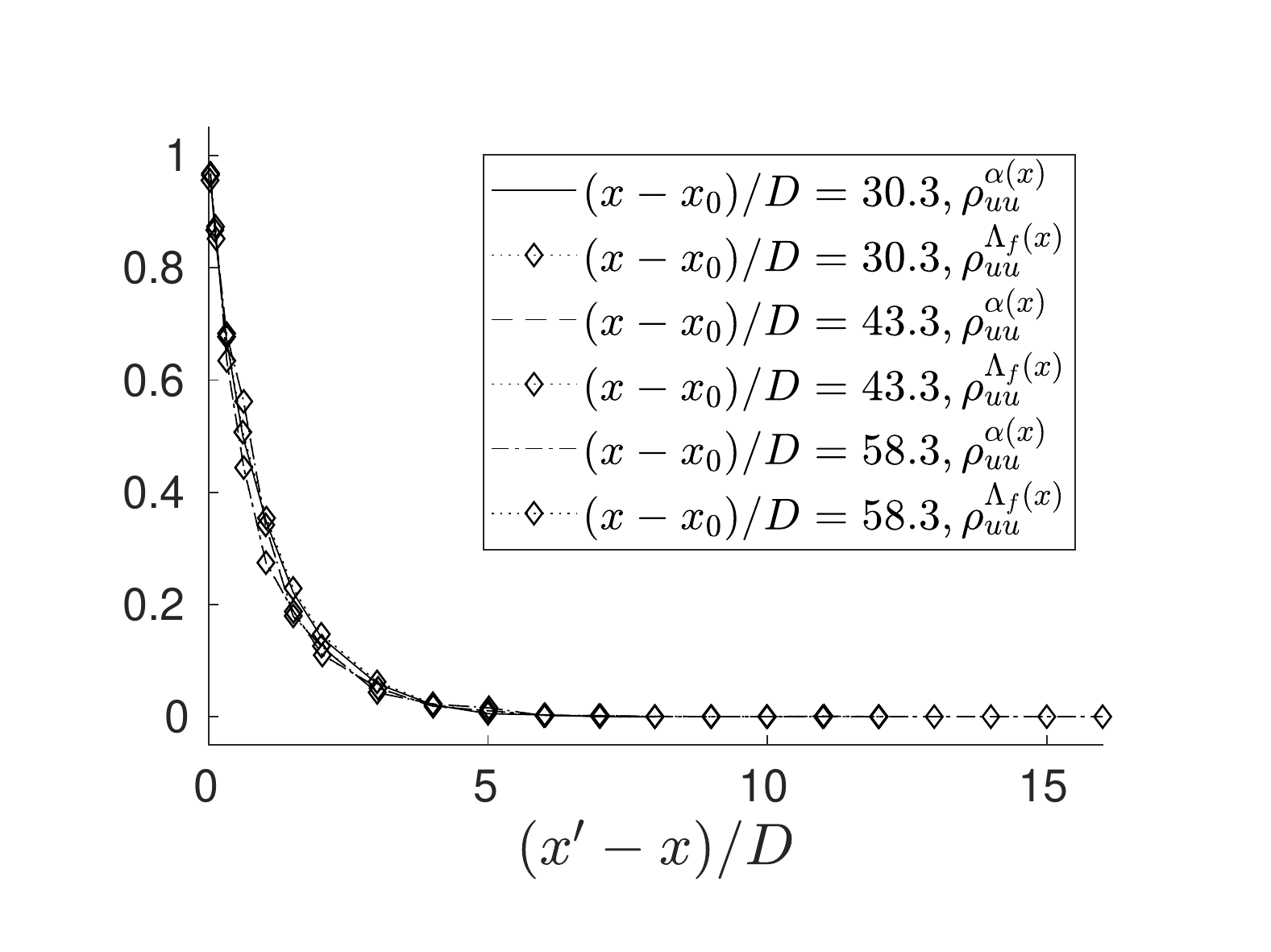}\label{fig:multiple_collapse}}
\subfloat[]{\includegraphics[scale=0.45]{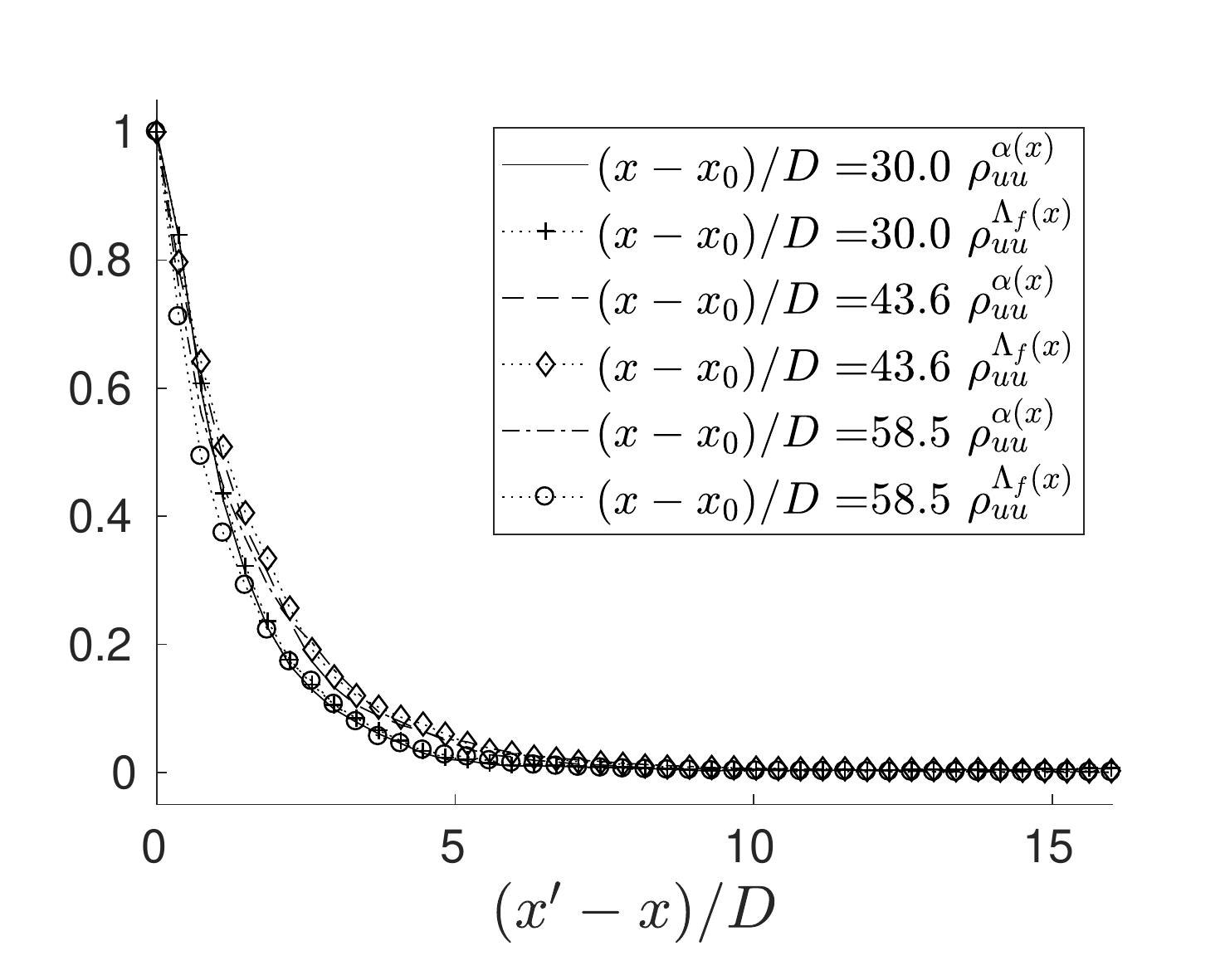}\label{fig:correlationtensor_new_alpha}}
\caption{(a): $\Lambda_f$, and $\alpha$-coefficients together with a linear curve fit of $\Lambda_f$ using data from \cite{Ewing2007}, and (b): corresponding results from current PIV data, (c): Demonstration of the collapse of ${\rho_{uu}^{\alpha(x)}}$ and $\rho_{uu}^{\Lambda_f(x)}$ using data from \cite{Ewing2007}, (d): corresponding results using current PIV data.} 
\end{figure}
\end{center}

The autocorrelation function is in essence the foundation for determining the integral scale. Since the general shape of the curves for different offset positions follow \eqref{eq:analytical_form}, the connection between $\alpha(x)$ and the integral scale is sought. The integral scale is commonly defined as 
\begin{equation}
\Lambda_f(x) = \int_0^\infty\rho_{uu}\left(x,x'\right)dx',
\end{equation}
such that the integral length scales are a function of the offset position, $x$. The estimates of the integral length scales are shown in figures \ref{fig:Lambda_alpha} and \ref{fig:Lambda_alpha_new} for the data of \cite{Ewing2007} and the PIV data (at the centerline), respectively, together with a linear fit. By direct comparison of the linear fits for $\alpha(x)$ and $\Lambda_f(x)$ it is observed that
\begin{equation}
\alpha(x) \approx \Lambda_f(x).\label{eq:alpha_equals_lambda}
\end{equation}
The validity of \eqref{eq:alpha_equals_lambda} is evaluated by comparing the measured $\Lambda_f$-values and $\alpha$-coefficients at each downstream position. The results in figures \ref{fig:Lambda_alpha} and \ref{fig:Lambda_alpha_new} are seen to be in good agreement with \eqref{eq:alpha_equals_lambda} for all downstream distances for both sets of data. These results indicate that $\alpha(x)$ is merely an expression of the integral length scale. It is clear from \eqref{eq:exp_alpha} that a displacement invariant form can be obtained using a power operation on the correlation coefficient.
\FloatBarrier
\noindent
This means that the displacement invariant form can be obtained from the two-point correlation function by the simple relation
\begin{equation}
e^{-|x'-x|} \approx \left(\frac{\overline{u(x,t)u(x',t)}}{a^{1,1}_c(*) U_s(x)U_s(x')}\right)^{\alpha(x)}\approx \left(\frac{\overline{u(x,t)u(x',t)}}{a^{1,1}_c(*) U_s(x)U_s(x')}\right)^{\Lambda_f(x)}.\label{eq:disp_invariant}
\end{equation}
This is confirmed by figure \ref{fig:multiple_collapse} showing a collapse of the two-point correlation tensor using both power operations, $(\cdot)^{\alpha(x)}$, and $(\cdot)^{\Lambda(x)}$. It is worth underlining here that a collapse of the two-point correlation was here obtained using a power-law operation in equidistant coordinates along the streamwise direction, without the strict necessity to transform to the logarithmically stretched coordinates, \eqref{eq:xi}.
\subsection{Consequences for the energy optimized basis}
Utilizing \eqref{eq:alpha_equals_lambda} the analytical form, \eqref{eq:analytical_form}, can be rewritten in terms of the integral length scale
\begin{equation}
\phi(x,x') = e^{-\frac{|x'-x|}{\alpha}} \approx e^{-sgn(x'-x)\frac{(x'-x)}{\Lambda_{f}}},
\end{equation}
where the norm in the last expression was replaced by the sign function. Neglecting the linear offset, $b$, in the linear representation of $\Lambda_f(x)$ yields
\begin{equation}
\phi(x,x') \approx e^{-sgn(x'-x)\frac{(x'-x)}{ax}}.
\end{equation}
Introducing the logarithmically stretched coordinates \eqref{eq:xi}, and \eqref{eq:upsilon} the expression can be written as
\begin{equation}
\phi(x,x') \approx e^{\frac{-sgn\left(e^{\xi'}-e^{\xi}\right)}{a}\left(e^{\upsilon}-1\right)},\label{eq:form}
\end{equation}
which is only a function of the logarithmic separation, $\upsilon = \xi'-\xi$, as predicted by \cite{Ewing2007}. Having concluded that the correlation coefficient is displacement invariant we revisit the conclusion from \cite{Ewing2007} that the energy-optimal way of decomposing the velocity field along the streamwise direction is by Fourier modes in similarity coordinates. 

The notion of Fourier modes being the optimal basis is founded on the idea that homogeneity has been achieved using the logarithmic coordinate transformation \eqref{eq:xi}, \cite{Ewing2007}. In the following the problem is re-investigated using the POD integral, in order to determine whether or not the decomposition can be performed using a Fourier transform in the jet far-field. The spatial POD integral takes the form
\begin{equation}
\int_\Omega R_{ij}(\textbf{x},\textbf{x'})\varphi_j(\textbf{x'})d\textbf{x'} = \lambda \varphi_i(\textbf{x}),\label{eq:POD}
\end{equation}
where the two-point correlation function is given by
\begin{equation}
R_{ij}(\textbf{x},\textbf{x'}) = \overline{u_i(\textbf{x},t)u'_j(\textbf{x'},t)}. \label{eq:correlation_function}
\end{equation}
\begin{center}
\begin{figure}[t]
\subfloat[]{\includegraphics[scale=0.45]{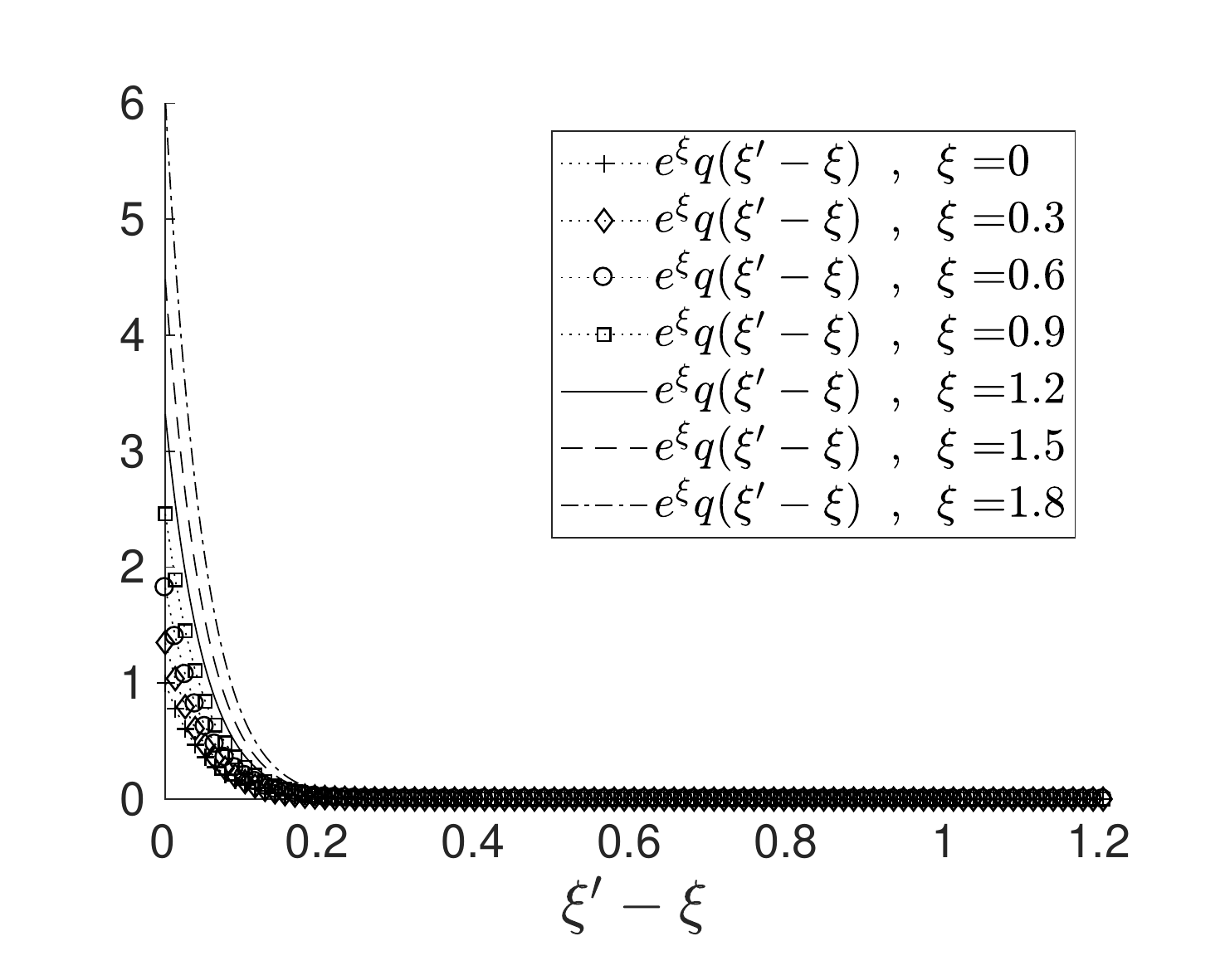}\label{fig:correlationtensor_new_exp_xi}}
\subfloat[]{\includegraphics[scale=0.45]{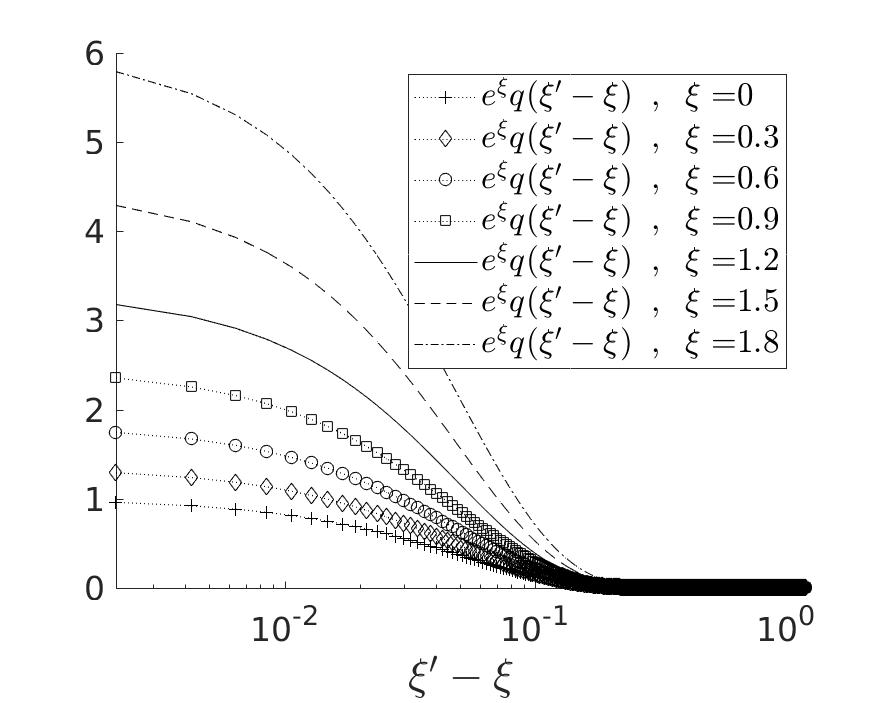}\label{fig:correlationtensor_new_exp_xi_log}}
\caption{The development of the POD kernel with downstream position along the centerline. (a): linear representation, (b): semi-log representation.}
\end{figure}
\end{center}
\noindent
We can relate \eqref{eq:correlation_coefficients} to \eqref{eq:correlation_function} by $R_{ij} = \rho_{ij}\sqrt{\overline{u_i^2}\,\overline{u_j'^2}}$. For the sake of the argument it is sufficient to restrict the analysis to the streamwise components, so $\rho_{uu}(\textbf{x},\textbf{x'})$ corresponds to $\rho_{ij}(\textbf{x},\textbf{x'})$ for $i=j=1$. If the streamwise Fourier basis is deductible from the POD, \eqref{eq:POD} must be characterized by a displacement kernel in that direction. In order to investigate this it is adequate to analyze the product of the correlation tensor and the Jacobian - here designated as $R_{ij}|J'|$ - as this term is required to be displacement invariant in order for the expansion using the Fourier basis to be applicable, \cite{lumley1967structure}. Note that this is most clearly seen by considering the ideal case of homogeneous turbulence in Cartesian coordinates, in which case the Jacobian is unity and the two-point correlation function is displacement invariant along the homogeneous direction, \cite{lumley1967structure}. The product, $R_{ij}|J'|$, therefore takes into account the volume element and hence the geometry of the chosen coordinate system. In this way it can be used to compare the jet far-field to a homogeneous turbulence field.

For the stretched similarity coordinates we find that $|J'| = A^2D^3e^{3\xi'}\eta'$, which means that the variation of the product of the two-point correlation tensor and the Jacobian along the streamwise direction for the streamwise component takes the following form
\begin{equation}
R_{11}(\xi',\xi)|J'| \approx a_c^{1,1}\underbrace{B^2M_0D^{-2}e^{-\xi}e^{-\xi'}}_{\propto U_s(\xi)U_s(\xi')}\underbrace{e^{\frac{-sgn\left(e^{\xi'}-e^{\xi}\right)}{a}\left(e^{\xi'-\xi}-1\right)}}_{h\left(\xi'-\xi\right)}\underbrace{A^2D^3e^{3\xi'}\eta'}_{|J'|},
\end{equation}
since $U_c(x) = BM^{1/2}_0/(x-x_0)$, \cite{Hussein1994}. This reduces to
\begin{equation}
R_{11}(\xi',\xi)|J'|\sim e^{\xi}\underbrace{e^{2\left(\xi'-\xi\right)}h\left(\xi'-\xi\right)}_{q\left(\xi'-\xi\right)} = e^{\xi}q\left(\upsilon\right).\label{eq:two-point_correlation}
\end{equation}
From \eqref{eq:two-point_correlation} it is seen that the kernel is a product of a displacement invariant part, ${q(\upsilon)=q\left(\xi'-\xi\right)}$, and a modulation function, $e^{\xi}$. Furthermore, since $q\left(\xi'-\xi\right)$ equals unity for $\xi'=\xi$ (or $\upsilon = 0$), it reveals that $e^{\xi}$ effectively increases the amplitude of two-point correlation function with downstream distance. This is demonstrated in figures \ref{fig:correlationtensor_new_exp_xi} and \ref{fig:correlationtensor_new_exp_xi_log} illustrating the streamwise growth of the kernel with downstream distance. These results show that 1) the kernel is not a displacement kernel, due to the presence of the modulation function, 2) far-field jet turbulence is fundamentally different from homogeneous turbulence, and 3) it cannot be directly concluded that the energy-optimized eigenfunctions in the streamwise direction are Fourier bases as otherwise stated in \cite{Ewing2007}. 
\FloatBarrier
\section{Conclusions}
Using two independent sets of experimental data sampled in the far-field region of the jet, the current work has demonstrated that the collapse of the two-point correlation coefficient along the streamwise direction in the jet far-field does not imply a displacement invariant POD kernel. The POD kernel in three dimensional space consists of the product of the two-point correlation tensor and the Jacobian, which was shown to be dependent on the absolute position in the far-field. This was concluded based on the exponentially increasing kernel with downstream distance. This result has the direct consequence that the jet far-field turbulence cannot be considered homogeneous along the streamwise direction in similarity coordinates, as the POD kernel in any homogeneous direction is characterized by displacement invariance.
\section*{Acknowledgments}
This project has received funding from the European Research Council (ERC) under the European Union’s Horizon 2020 research and innovation program (grant agreement No 803419).
\section*{Declaration of interests}
The authors report no conflict of interest.
\bibliographystyle{jfm}


%
\end{document}